\documentclass[floats,floatfix,showpacs,amssymb,prd,twocolumn,superscriptaddress,nofootinbib,nolongbibliography,reprint]{revtex4-2}

\usepackage{amssymb,amsmath,verbatim,mathtools,needspace,enumerate,etoolbox,graphicx,physics,microtype,afterpage,bigints,gensymb,tabularx,xspace,tensor,braket}

\usepackage[dvipsnames, usenames]{xcolor}
\definecolor{linkcolor}{rgb}{0.0,0.3,0.5}
\definecolor{dodgerblue}{HTML}{1E90FF}
\usepackage[unicode, colorlinks=true, linkcolor=linkcolor, citecolor=linkcolor, filecolor=linkcolor,urlcolor=linkcolor, pdfusetitle]{hyperref}
\usepackage[all]{hypcap}
\usepackage[T1]{fontenc}
\usepackage[utf8]{inputenc}
\usepackage{orcidlink}
\usepackage{fontawesome}

\graphicspath{{fig}}

\newcommand{\bham}{\affiliation{School of Physics and Astronomy \& Institute for Gravitational Wave Astronomy, University of Birmingham, \\ Birmingham, B15 2TT, United Kingdom}}
\newcommand{\milan}{\affiliation{Dipartimento di Fisica ``G. Occhialini'', 
Universit\'a degli Studi di Milano-Bicocca, Piazza della Scienza 3, 20126 Milano, Italy}}
\newcommand{\infn}{\affiliation{INFN, Sezione di Milano-Bicocca, 
Piazza della Scienza 3, 20126 Milano, Italy}}
\newcommand{\como}{\affiliation{Dipartimento di Scienza e Alta Tecnologia, Universit\'a dell’Insubria, via Valleggio 11, 22100 Como, Italy}}

\date{\today}

\begin{document}
\title{
Flexible mapping of ringdown amplitudes for nonprecessing binary black holes
}

\author{Costantino Pacilio$\,$\orcidlink{0000-0002-8140-4992}}
\email{costantino.pacilio@unimib.it}
\milan \infn 

\author{Swetha Bhagwat$\,$\orcidlink{0000-0003-4700-5274}}
\email{s.bhagwat@bham.ac.uk}
\bham

\author{Francesco Nobili$\,$\orcidlink{0009-0008-8769-4557}}
\email{fnobili@uninsubria.it}
\como  \infn

\author{Davide Gerosa$\,$\orcidlink{0000-0002-0933-3579}}
\email{davide.gerosa@unimib.it}
\milan \infn

\begin{abstract}
The remnant black hole from a binary coalescence emits ringdown gravitational waves characterized by quasinormal modes, which depend solely on the remnant's mass and spin. In contrast, the ringdown amplitudes and phases are determined by the properties of the merging progenitors. Accurately modeling these amplitudes and phases reduces systematic biases in parameter estimation and enables the development and performance of rigorous tests of general relativity. We present a state-of-the-art, data-driven surrogate model for ringdown amplitudes and phases, leveraging Gaussian process regression trained against SXS numerical-relativity simulations. Focusing on nonprecessing, quasicircular binary black holes, our model offers the most comprehensive fit that includes 16 emission modes, incorporating overtones and quadratic contributions. Our surrogate model achieves reconstruction errors that are approximately 2 orders of magnitude smaller than the typical measurement errors of current gravitational-wave interferometers.  An additional benefit of our approach is its flexibility, which allows for future extensions to include features such as eccentricity and precession, broadening the scope of its applicability to more generic astrophysical scenarios. Finally, we are releasing our model in a ready-to-use package called~\texttt{postmerger}. 
\href{https://github.com/cpacilio/postmerger}{\large\faGithub}
\end{abstract}

\maketitle
\section{Introduction}
\label{sec:intro}
Uniqueness theorems \cite{Israel:1967wq,Hawking:1971vc,Carter:1971zc,Robinson:1975bv} in general relativity (GR) state that stationary, asymptotically flat black-hole (BH) spacetimes are described by the Kerr metric \cite{Kerr:1963ud,Teukolsky:2014vca}, which is uniquely determined by a mass and a spin parameter. A more general solution including electric and/or magnetic charge is also allowed in theory \cite{Newman:1965my}, but astrophysical BHs are considered neutral due to discharging effects \cite{Gibbons:1975kk,Goldreich:1969sb,1975ApJ...196...51R,1977MNRAS.179..433B}. This statement is often referred to as the ``no-hair'' theorem and holds for BH perturbations as well. Numerical solutions to the Teukolsky's equations  \cite{Teukolsky:1972my,Teukolsky:1973ha} show that perturbed BHs relax toward equilibrium by shedding higher multipole moments through the emission of gravitational waves (GWs), also known as ringdown. In particular, ringdown GWs are radiated away at characteristic complex frequencies~\cite{Vishveshwara:1970zz,Press:1971wr,Press:1973zz}, or quasinormal modes (QNMs).

The QNM spectrum of binary BHs with astrophysical masses \cite{Kokkotas:1999bd,Berti:2009kk} is such that their ringdown can be detected by current and future GW detectors \cite{Berti:2016lat}. Crucially, detecting multiple QNMs ---also known as BH spectroscopy~\cite{Dreyer:2003bv}--- allows us to perform tests of GR in the strong-field regime~\cite{Berti:2005ys,Berti:2007zu,Meidam:2014jpa,Berti:2018vdi}. Some non-GR theories of gravity predict deviations from the no-hair theorem and these are reflected in the QNM frequencies of ringing BHs~\cite{Barausse:2008xv,Berti:2018vdi,Carullo:2021oxn,Franchini:2023eda}. BH spectroscopy with the current LIGO/Virgo detectors is feasible \cite{Carullo:2019flw,Capano:2021etf,Ghosh:2021mrv,Siegel:2023lxl,LIGOScientific:2020tif,LIGOScientific:2021sio} but challenging due to the relatively low signal-to-noise ratio in the ringdowns. 
 Next-generation GW detectors will be able to test GR from ringdown QNM spectra with percent-level precision~\cite{Berti:2016lat,Bhagwat:2023jwv,Pacilio:2023mvk}. 
 These efforts require the use of ringdown waveform templates which are accurate and include all known physical effects to reduce systematic biases \cite{Toubiana:2023cwr,Toubiana:2024car,Gupta:2024gun}. The most accurate approach toward solving the two-body problem in GR is numerical relativity (NR), but simulations are expensive and cannot be used on the fly in GW parameter-estimation studies.

We plan to build a flexible, data-driven model for the amplitudes of excitation of QNMs in ringdown waveforms by directly interpolating a training bank of NR simulations. This is an ambitious program and, as a first step in this paper, we focus on nonprecessing quasicircular binary BH systems. Amplitude excitations of various sets of QNMs for these sources have also been tackled by several authors \cite{Kamaretsos:2012bs,London:2014cma,Baibhav:2017jhs,London:2018gaq,Borhanian:2019kxt,Forteza:2022tgq,Cheung:2023vki}.
In particular, Ref.~\cite{London:2018gaq} proposed closed-form expressions for the ringdown excitation amplitudes of multiple ringdown fundamental tones. This model includes harmonics that are accessible with the sensitivities of current GW detectors \cite{Cotesta:2018fcv,Pompili:2023tna} and it has since been used 
to perform NR-informed parameter estimations of BH ringdowns by the LVK Collaboration \cite{LIGOScientific:2020tif,LIGOScientific:2021sio}. Knowledge of the excitation amplitudes of perturbed BHs is also important to enrich the scope of ringdown tests of GR beyond BH spectroscopy: for example, one can perform consistency checks between the amplitudes as measured by assuming agnostic ringdown templates and those predicted by fitting expressions calibrated to GR~\cite{Bhagwat:2021kfa,Forteza:2022tgq}.

Instead of assuming closed-form ansatzes, here we take a different approach to ringdown waveform modeling by turning to data-driven fits, specifically to Gaussian process regression (GPR). 
GPR is a flavor of nonparametric algorithms for supervised learning \cite{williams2006gaussian,mackay2003information,2023CSE....25d...4W} and produces a probability distribution of best-fit functions: 
more precisely, it models the output quantities as random variables with joint Gaussian distributions, 
such that the mean and covariance depend on the input quantities. GPR does not require a prior ansatz of the functional dependence of the outputs 
and, unlike most of the machine-learning regression strategies on the market, comes with an estimate of prediction errors across the parameter space. This is a very desirable feature in our context as it allows users to assess the confidence of the predictions and plan for new set of training simulations~\cite{2023PhRvD.107b4034F}. 
In BH binary dynamics, GPR was used to build surrogate models of the BH remnant properties as function of the initial binary parameters \cite{Varma:2018aht,Boschini:2023ryi}. 

In this work, we train a GPR model on the set of nonprecessing (both coaligned and counteraligned) NR waveforms made publicly available in the Simulating eXtreme Spacetimes (SXS) catalog \cite{Boyle:2019kee}. We assess the goodness of our fits both internally and by comparing with existing parametric fits such as those of Refs.~\cite{London:2018gaq,Cheung:2023vki}. Additionally, our fits incorporate contributions from the first overtones, which are missing from the model of Ref.~\cite{London:2018gaq}.
When comparing against results from Ref.~\cite{Cheung:2023vki}, we find that the additional flexibility of our models translates into more accurate predictions.
Furthermore, we also model contributions from nonlinear quadratic modes within some of the 
 harmonics. Our surrogate models are ready to be used for GW parameter estimation. The GPR uncertainties can be used to amortize systematic biases in precision tests of GR.

The work is organized as follows.  In Sec.~\ref{sec:notation} we introduce our conventions.  
In Sec.~\ref{sec:catalog} we detail the extraction of ringdown data from the NR waveforms. In Sec.~\ref{sec:gpr} we describe our GPR model, including the choice of fitting variables and the training strategy. In Sec.~\ref{sec:results} we present our predictions and compare results against previous findings. A discussion and outlook are presented in Sec.~\ref{sec:conclusions}.
\section{Ringdown waveforms}
\label{sec:notation}
The most general expression for the ringdown waveform in GR is given by 
\begin{equation}
    \label{eq:strain:1}
    h(t)=\sum_{l,m,n}\mathcal{A}_{lmn}e^{-i\omega_{lmn}(t-t_0)}\tensor[_{-2}]{S}{_{lmn}}(\iota,\varphi,M\chi\omega_{lmn})
\end{equation}
where
\begin{enumerate}[(i)]
    \item $l=2,3,\dots$ is the angular index, $m=-l,\dots,+l$ is the azimuthal index, and $n=0,1,\dots$ is the overtone index;
    \item 
    $\omega_{lmn}=2\pi f_{lmn}-i/\tau_{lmn}$ are the complex QNM frequencies of a ringing BH with mass $M$ and spin~$\chi$;
    \item $\mathcal{A}_{lmn}$ are the complex amplitudes corresponding to the $(l,m,n)$ excitations;
    \item $\tensor[_{-2}]{S}{_{lmn}}(\iota,\varphi,M\chi\omega_{lmn})$ are the spin-2 \emph{spheroidal} harmonics; 
    \item $t_0$ is a conventional starting time for the ringdown portion of the waveform.
\end{enumerate}

The GW strain in NR simulations is usually expanded in spin-2 \textit{spherical} harmonics as
\begin{equation}
    \label{eq:strain:2}
    \begin{split}
    h(t) &= h_+(t) - i h_\times(t) = \sum_{l,m}h_{lm}(t)\;\;\tensor[_{-2}]{Y}{_{lm}}(\iota,\varphi)\,.
    \end{split}
\end{equation}
Expressions \eqref{eq:strain:1} and \eqref{eq:strain:2} are equivalent upon expanding the spheroidal harmonics in terms of the spherical harmonics 
\begin{equation}
    \label{eq:mixing:1}
    \tensor[_{-2}]{S}{_{lmn}}(\theta,\varphi,M\chi\omega_{lmn}) = \sum_{l'} \mu^*_{ml'ln} \;\; \tensor[_{-2}]{Y}{_{l'm}}(\theta,\varphi)
\end{equation}
where $\mu_{mll'n}$ are complex numbers in the notation of Ref.~\cite{Berti:2014fga} and $\,^*$ indicates complex conjugation. One can write the $h_{lm}$ components of the ringdown strain as
\begin{equation}
    \label{eq:strain:3}
    h_{lm}(t)=\sum_{l',n}\mu^*_{mll'n}\mathcal{A}_{l'mn}e^{-i\omega_{l'mn}(t-t_0)}\,.
\end{equation}
Each \textit{spherical} component $(l,m)$ corresponds to a superposition of  (\emph{spheroidal}) QNMs with the same value of $m$ but different values of $l$. 

While Eq.~\eqref{eq:strain:3} is generic, in practice one has that the value of $\mu_{mll'n}$ for $l'\in\{l,l-1\}$ is much greater than that of all the other cases  \cite{Berti:2014fga}, which can thus be neglected~\cite{KumarMehta:2019izs,Pompili:2023tna}.
With this approximation one has for example that
\begin{equation}
\begin{split}
    h_{32}(t)\approx&\sum_{n}\left[\mu^*_{233n}\mathcal{A}_{32n}e^{-i\omega_{32n}(t-t_0)}\right.\\
    &\left.+\mu^*_{232n}\mathcal{A}_{22n}e^{-i\omega_{22n}(t-t_0)}\right]\,.
\end{split}
\end{equation}
Finally, it is convenient to introduce the polar decomposition
\begin{equation}
    \mathcal{A}_{lmn}=A_{lmn}e^{-i\phi_{lmn}}
\end{equation}
where $A_{lmn}$ and $\phi_{lmn}$ are real valued.
We evaluate the QNM spectrum of a Kerr BH from Eqs.~\eqref{eq:strain:1}-\eqref{eq:strain:3} by a quadratic interpolation of the numerical data provided with Ref.~\cite{Berti:2005ys}. 

Note that in Eq.~\eqref{eq:strain:1} and throughout this work we neglect the contribution from the retrograde modes to the ringdown. Different conventions have been used on the definition of QNMs, including the definition of retrograde modes and their relation to negative-$m$ modes, as well as the definition of QNMs for negative BH spins. We adopt the same conventions of \cite{Cheung:2023vki} allowing for a unified treatment of positive and negative spins, as opposed to the conventions in \cite{MaganaZertuche:2021syq} requiring to switch from prograde to retrograde modes when turning to negative spins. This choice allows us to smoothly interpolate ringdown amplitudes across the parameter space, without introducing \textit{ad hoc} prograde-retrograde switches. We further clarify our conventions in the Appendix \ref{sec:qnms}. 
\section{Numerical multimode extraction}
\label{sec:catalog}
We extract numerical ringdown waveforms from the public release of the SXS catalog \cite{Boyle:2019kee} containing binary BH simulations labeled from \texttt{BBH:0001} to \texttt{BBH:2265}. 
We use extrapolated waveforms in the outermost extrapolation order, corrected with the average motion of the system's center of mass~\cite{Boyle:2015nqa}.
We restrict our training set to quasicircular, nonprecessing binaries. In particular, 
we selected sources with reported eccentricity $e<10^{-3}$ and 
in plane spin components
$|\chi_{x,y}|/||\vec{\chi}||<10^{-3}$ or $||\vec\chi||<10^{-3}$ 
for each progenitor spin. The latter conditions can be violated because of numerical artifacts at small spin magnitudes; therefore, we consider progenitors with spin magnitudes $||\vec{\chi}||<10^{-3}$ as ``{nonspinning}'' even when they violate the above in-plane spin constraints. This selection results in a total of 437 numerical simulations.

Isolating the ringdown portion within a NR waveform is challenging, because the transition from the merger to the ringdown is not sharp \cite{Bhagwat:2017tkm}. A rule of thumb is that the system enters the ringdown regime at $t_{\rm start}\approx 20M_{\rm i}$  
after the peak of $||h_{22}||^2$ \cite{Kamaretsos:2012bs,London:2014cma,London:2018gaq}, 
where $M_{\rm i}$ is the initial mass of the binary. 
Less conservative choices such as $t_{\rm start}\sim10M_{\rm i}$ or $t_{\rm start}\sim15M_{\rm i}$ induce systematic biases that are below the current data's statistical uncertainties \cite{Bhagwat:2017tkm,Carullo:2018sfu}
and have been adopted to maximize the signal-to-noise ratio of current ringdown data \cite{Carullo:2019flw,Forteza:2022tgq,Capano:2021etf,LIGOScientific:2020tif,LIGOScientific:2021sio}. References~\cite{London:2018gaq,Baibhav:2023clw,Cheung:2023vki} propose the more rigorous criterion that the amplitudes $\mathcal{A}_{lmn}$ 
are constant when extracted at different starting times, as implied by Eq.~\eqref{eq:strain:1}. This generally results in different starting times for each mode, which considerably complicates the analysis \cite{Cheung:2023vki}. 
Here, we do not attempt at contributing to this discussion and simply fix the starting time at the conservative value of $20 M_{\rm i}$ after the peak of $||h_{22}||^2$. Moreover, we truncate all waveforms at $t_{\rm end}=t_{\rm peak}+100M_{\rm i}$ after the peak to exclude spurious features due to numerical noise.

The choice of the starting time impacts the multipolar content of the signal.
Previous multipolar fits of numerical ringdown waveforms show that, for the choice of $t_{\rm start}=t_{\rm peak}+20M_{\rm i}$, multiple modes confidently stand above numerical noise and their functional dependence on the initial parameters can be explored \cite{Kamaretsos:2012bs,London:2014cma,London:2018nxs,Forteza:2022tgq,Cheung:2023vki}. Following Refs.~\cite{London:2018gaq,Pompili:2023tna,Cheung:2023vki}, we thus include  modes with $(l,m)=$ (2,2), (2,1), (3,3), (3,2), (4,3), (4,4) and (5,5). For all of these, we consider both the fundamental modes ($n=0$) as well as the fist overtone ($n=1$). In data analysis, the omission of overtones  
can lead to biased estimates of the mass and spin of the ringing BH in the high SNR limit \cite{Baibhav:2017jhs,Giesler:2019uxc} 
while their inclusion 
stabilizes the recovery of the amplitudes and phases of the fundamental $n=0$ modes \cite{Baibhav:2017jhs,London:2018gaq,JimenezForteza:2020cve,Clarke:2024lwi}. At present, it is unclear whether overtones are physically present or simply alleviate the fitting residuals \cite{Baibhav:2023clw,Nee:2023osy}. 
Finally, numerous studies have highlighted the importance of nonlinear modes in modelling some of the harmonics \cite{London:2014cma,Mitman:2022qdl,Cheung:2022rbm}. Here, we also include the quadratic $(2,2,0)\times(2,2,0)$ and $(2,2,0)\times(3,3,0)$ modes when fitting the $(4,4)$ and $(5,5)$ harmonics respectively. The modes captured in this study are listed in Table~\ref{tab:modes}. 

\begin{table}[]
    \centering
    \begin{tabular}{c@{\hskip 10pt}|@{\hskip 10pt}c}
    $(l,|m|)$ & Contributing modes $(l,|m|,n)$ \\
    \hline
    \hline
    $(2,2)$ & $(2,2,0),~(2,2,1)$ \\
    $(2,1)$ & $(2,1,0),~(2,1,1)$\\
    $(3,3)$ & $(3,3,0),~(3,3,1)$\\
    $(3,2)$ & $(3,2,0),~(3,2,1),~(2,2,0)\dag,~(2,2,1)\dag$\\
    $(4,3)$ & $(4,3,0),~(4,3,1),~(3,3,0)\dag,~(3,3,1)\dag$\\
    $(4,4)$ & $(4,4,0),~(4,4,1),~(2,2,0)\times(2,2,0)$\\
    $(5,5)$ & $(5,5,0),~(5,5,1),~(2,2,0)\times(3,3,0)$\\
    \end{tabular}
    \caption{List of the (\textit{spherical}) harmonic indices $(l,|m|)$ and of the QNMs that we include within them in our surrogate model. The symbol $\times$ indicates quadratic coupling between modes. QNMs marked with the symbol $\dag$ are not fitted independently, but the spherical-spheroidal mixing relations \eqref{eq:strain:3} are assumed.
    }
     \label{tab:modes}
\end{table}
Note that, differently from Refs.~\cite{London:2018gaq,Cheung:2023vki}, we do not model independently the spherical-spheroidal mixing modes. Rather, we impose that they satisfy Eq.~\eqref{eq:strain:3} where the mixing coefficients $\mu_{mll'n}$ are interpolated from the numerical tables\footnote{We account for a correction factor $\mu_{mll'n}\to(-1)^{l+l'}\mu_{mll'n}$ due to different conventions in the definition of the spin-weighted spherical harmonics between Ref.~\cite{Berti:2014fga} and the SXS code \cite{KumarMehta:2019izs}.} of Ref.~\cite{Berti:2014fga}.

For nonprecessing binaries, equatorial symmetry implies $h_{l-m}=(-1)^{l}h^*_{lm}$ \cite{Borchers:2022pah}. Therefore, we do not model negative-$m$ modes as they are simply related to positive-$m$ modes. This is not true when the progenitor spins precede, but this is outside the scope of this work. 

We fit SXS waveforms against the functional form of Eq.~\eqref{eq:strain:3} using nonlinear least squares as implemented in the \texttt{scipy.curvefit}~\cite{2020SciPy-NMeth}. Specifically, we fix the frequencies and damping times to those of a Kerr BH with the final mass and final spin estimated from the metadata of the NR simulation and we leave $\{A_{lmn},\phi_{lmn}\}$ as free parameters, with the only exception of the spherical-spheroidal mixing modes, as explained above. We impose that the amplitudes $A_{lmn}$ are positive and search for the phases $\phi_{lmn}$ in the interval $[0,2\pi]$. We quantify the goodness of our fits by the mismatch 
\begin{equation}
    \label{eq:mismatch}
    \mathcal{M}_{lm}=1-\frac{\braket{h_{lm}^{\rm NR}|h_{lm}^{\rm fit}}}{\sqrt{\braket{h_{lm}^{\rm NR}|h_{lm}^{\rm NR}}\braket{h_{lm}^{\rm fit}|h_{lm}^{\rm fit}}}}
\end{equation}
where $h_{lm}^{\rm NR}$ are the NR waveforms and the superscript $^{\rm fit}$ indicates quantities fitted against NR data, and the inner product is defined as
\begin{equation}
    \label{eq:inner:product}
    \braket{a|b}={\rm Re}\left[\int_{t_{\rm start}}^{t_{\rm end}}dt~a(t)b^*(t)\right]\,.
\end{equation}
\section{GPR surrogates}
\label{sec:gpr}
\subsection{Input variables}
\label{sec:input}
Vacuum GR
is a scale-free theory which implies, nonprecessing binary BHs 
are characterized by three degrees of freedom: the mass ratio $q=m_1/m_2\geq1$ 
and the $z$-components of the spins $\chi_{1z}$ and $\chi_{2z}$, where the $z$ axis is aligned to the orbital angular momentum of the binary. References~\cite{Kamaretsos:2012bs,London:2018gaq,Forteza:2022tgq,Cheung:2023vki} look for convenient functional forms to interpolate the complex amplitudes $\mathcal{A}_{lmn}$, or equivalently the real valued amplitudes $A_{lmn}$ and phases $\phi_{lmn}$, in terms of the initial binary parameters. 
In all cases, educated guesses are made about which combinations of the original independent variables $\{q,\chi_{1z},\chi_{2z}\}$  better capture trends in the data, finding that  $\eta=q/(1+q)^2$ and $\delta=\sqrt{1-4\eta}$ result in more accurate fits compared to the mass ratio $q$, while the symmetric and antisymmetric spins $\chi_s=(q\chi_{1z}+\chi_{2z})/(1+q)$ and $\chi_a=(q\chi_{1z}-\chi_{2z})/(1+q)$ are more suitable than the individual spins $\chi_{1z}$ and $\chi_{2z}$.\footnote{In GW astronomy, $\chi_s$ is also referred to as the effective spin $\chi_{\rm eff}$~\cite{Damour:2001tu}.} 

When fitting the NR data, these parameters are usually combined in \textit{ad hoc} functional forms. For example, Ref.~\cite{London:2018gaq} uses complex-valued polynomials for the complex amplitudes $\mathcal{A}_{lmn}$, while \cite{Forteza:2022tgq,Cheung:2022rbm} use real-valued polynomials to fit the amplitudes $A_{lmn}$ and phases $\phi_{lmn}$ separately. However, these functional forms are subjected to a certain arbitrariness (e.g., the degree of the polynomial expansion) which are resolved on a case-by-case basis, and they do not necessarily behave nicely outside the calibration region. Our approach is instead that of using a nonparametric strategy, specifically GPR.

We build GPR fits that map the independent variables $\{\delta,\chi_s,\chi_a\}$ into the complex amplitudes of excitation of each ringdown mode. 
After investigating $q$, $\eta$ and $\log q$ as possible alternatives, we found that using $\delta$ gives better interpolation results.
\subsection{Output variables}
\label{sec:output}
The raw phases $\phi_{lmn}$ extracted from the NR simulations depend on the initial conditions, only their difference is meaningful. In other words, one of the phases, say $\phi_{220}$, can be reabsorbed within  the azimuthal angle $\varphi$ entering the spherical harmonics. We construct physically meaningful phases $\tilde\phi_{lmn}$ relative to $(2,2,0)$ by performing an azimuthal rotation that conventionally sends $\phi_{220}$ to 0, i.e.
\begin{equation}
    \tilde\phi_{lmn}=\phi_{lmn}-\frac{m}{2}\phi_{220}\,.
\end{equation}
 
NR catalogs adopt phase conventions such that fitted points in the $(A_{lmn}-\tilde\phi_{lmn})$ plane repeat periodically along the $\tilde\phi_{lmn}$ direction, with periodicity ${\rm Per}(\tilde\phi)=\pi$ for odd-$m$ modes and ${\rm Per}(\tilde\phi)=2\pi$ for even-$m$ modes.
This induces discontinuities and redundancies in the $(A_{lmn}-\tilde\phi_{lmn})$  plane, which makes direct fits cumbersome. Previous attempts have dealt with this issue by fitting either ${\rm mod}[2\phi_{lmn},2\pi]$~\cite{Cheung:2023vki} or ${\rm mod}[\phi_{lmn},{\rm Per}(\tilde\phi)]$~\cite{Forteza:2022tgq}. In both approaches, one deals with discontinuities by ``unwrapping'' the phase, i.e., shifting certain values by integer multiples of either $2\pi$ or ${\rm Per}(\tilde\phi)$ to enforce continuity. As also observed in Ref.~\cite{Carullo:2024smg}, we found that this strategy is simple to implement for the $(2,2,n)$, $(2,1,n)$, $(3,3,n)$, $(3,2,0)$ and $(4,4,0)$ modes, but it does not seem to be practical in the general case. Alternatively, one can fit directly the complex amplitudes $\mathcal{A}_{lmn}$ \cite{London:2018gaq}.

Our strategy goes as follows. 
We introduce effective Cartesian coordinates 
\begin{align}
    \label{eq:cartesian:1}
    \tilde A_{lmn}^x&=
    \begin{cases}
        A_{220}&\text{for $(l,m,n)=(2,2,0)$,}\\
        A^{\rm R}_{lmn}\cos(\beta\tilde\phi_{lmn})&\text{\rm otherwise,}
    \end{cases}
\\
    \label{eq:cartesian:2}
    \tilde A_{lmn}^y&=
    \begin{cases}
        0&\text{for $(l,m,n)=(2,2,0)$,}\\
        A^{\rm R}_{lmn}\sin(\beta\tilde\phi_{lmn})&\text{\rm otherwise,}
    \end{cases}
\end{align}
where
\begin{equation}
    \label{eq:amp:relative}
    A^{\rm R}_{lmn}=\frac{A_{lmn}}{A_{220}}
\end{equation}
are the relative amplitudes to $(2,2,0)$ and the phase factor is $\beta=1$ for even-$m$ modes and $\beta=2$ for odd-$m$ modes. 

Using relative amplitudes to the $(2,2,0)$ mode is motivated by the fact that the latter it is generally the most excited mode in nonprecessing systems and $A_{220}$ has a simple profile \cite{London:2014cma,London:2018gaq}; indeed, previous studies~\cite{Kamaretsos:2012bs,Forteza:2022tgq} reported that relative amplitudes behave smoothly across the parameter space. Polar variables for $(l,m,n)\neq(2,2,0)$ are given by
\begin{equation}
    \label{eq:polar:1}
    A^{\rm R}_{lmn}={\rm abs}\left(\tilde A_{lmn}^x+i\tilde A_{lmn}^y\right)
\end{equation}
and
\begin{equation}
    \label{eq:polar:2}
    \tilde\phi_{lmn}=\frac{{\rm angle}\left(\tilde A_{lmn}^x+i\tilde A_{lmn}^y\right)}{\beta}\,.
\end{equation}
We use the \texttt{numpy.abs} and \texttt{numpy.angle} routines \cite{harris2020array}. 
The latter returns outputs with support in the interval $[-\pi,\pi]$: when comparing with existing fits from the literature, we adapt their  conventions for the domain of definition of the phase to that used here.

\begin{figure*}
    \centering
    \includegraphics[width=0.8\textwidth]{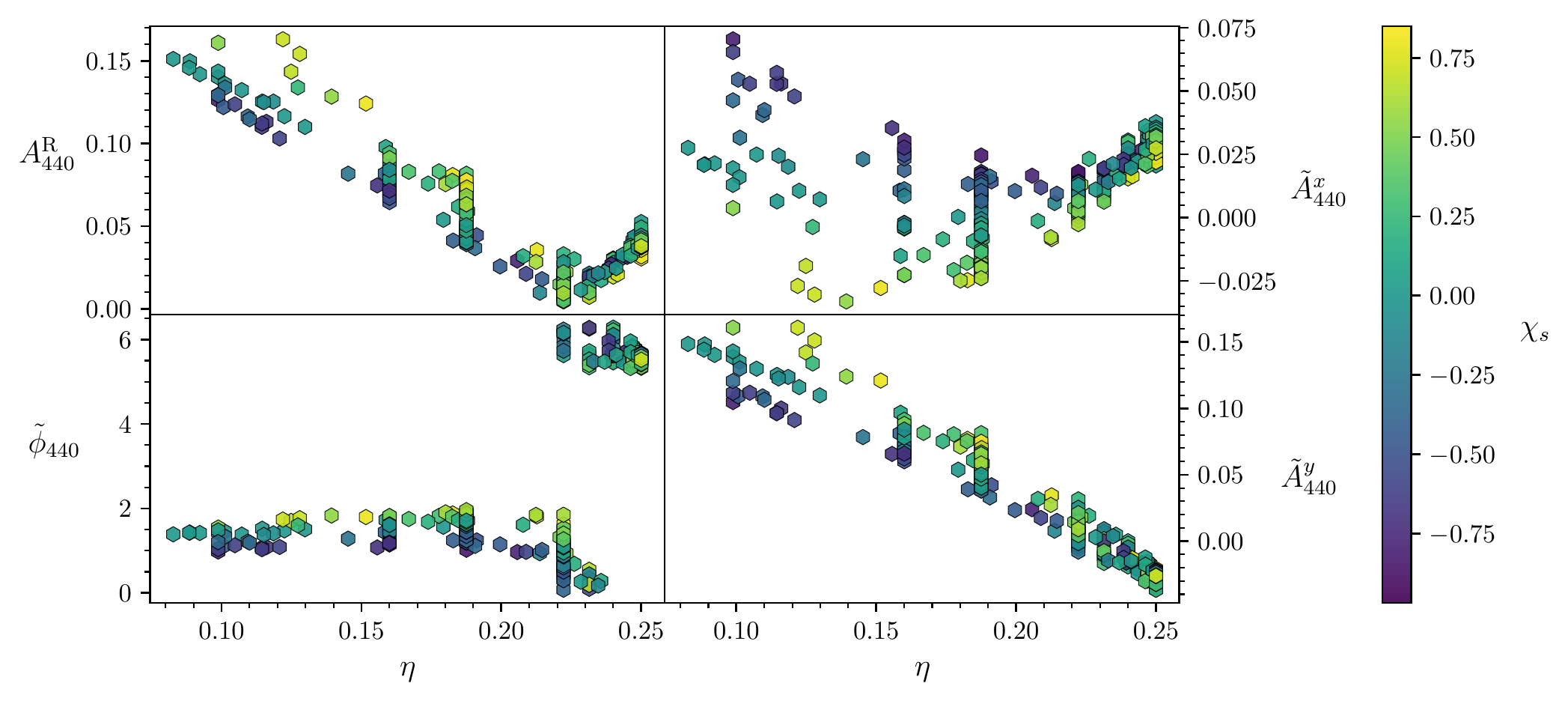}
    \caption{\textit{Left panels:} Relative amplitudes $A_{440}^{\rm R}$ and phases $\tilde\phi_{440}$ across the NR catalog used here, as a function of the symmetric mass ratio $\eta$ (horizontal axis) and of the symmetric spin $\chi_s$ (color bar). \textit{Right panels:} Same as left, but for the effective Cartesian components $A_{440}^x$ and $A_{440}^y$. In all subplots, we only plot the data points marked as \textit{inliers}, as defined in Sec.~\ref{sec:training:strategy}.}
    \label{fig:res:4}
\end{figure*}

The advantage of working with effective Cartesian coordinates to avoid discontinuities is illustrated in Fig.~\ref{fig:res:4}. The left panels show the relative amplitudes and phases of the $(4,4,0)$ mode across the NR dataset. The relative phases have been redefined in the interval $[0,2\pi]$ and present a discontinuity. Moreover, the profile of $A_{lmn}^{\rm R}$ has a nonsmooth cusplike behavior at symmetric mass ratio $\eta\approx0.23$. While in this case the discontinuity in the relative phases could be adjusted by hand with a simple shifting of the upper disconnected region by $-2\pi$, the right panels show that the effective Cartesian coordinates vary smoothly across the parameter space.\footnote{In earlier attempts during the preparation of this work, we considered using a Matern kernel \cite{williams2006gaussian} to produce a GPR surrogate directly for the discontinuous quantities $\tilde\phi_{lmn}$. However, we find that using effective Cartesian coordinates produces better results.}

\begin{figure*}
    \centering
    \includegraphics[width=0.825\textwidth]{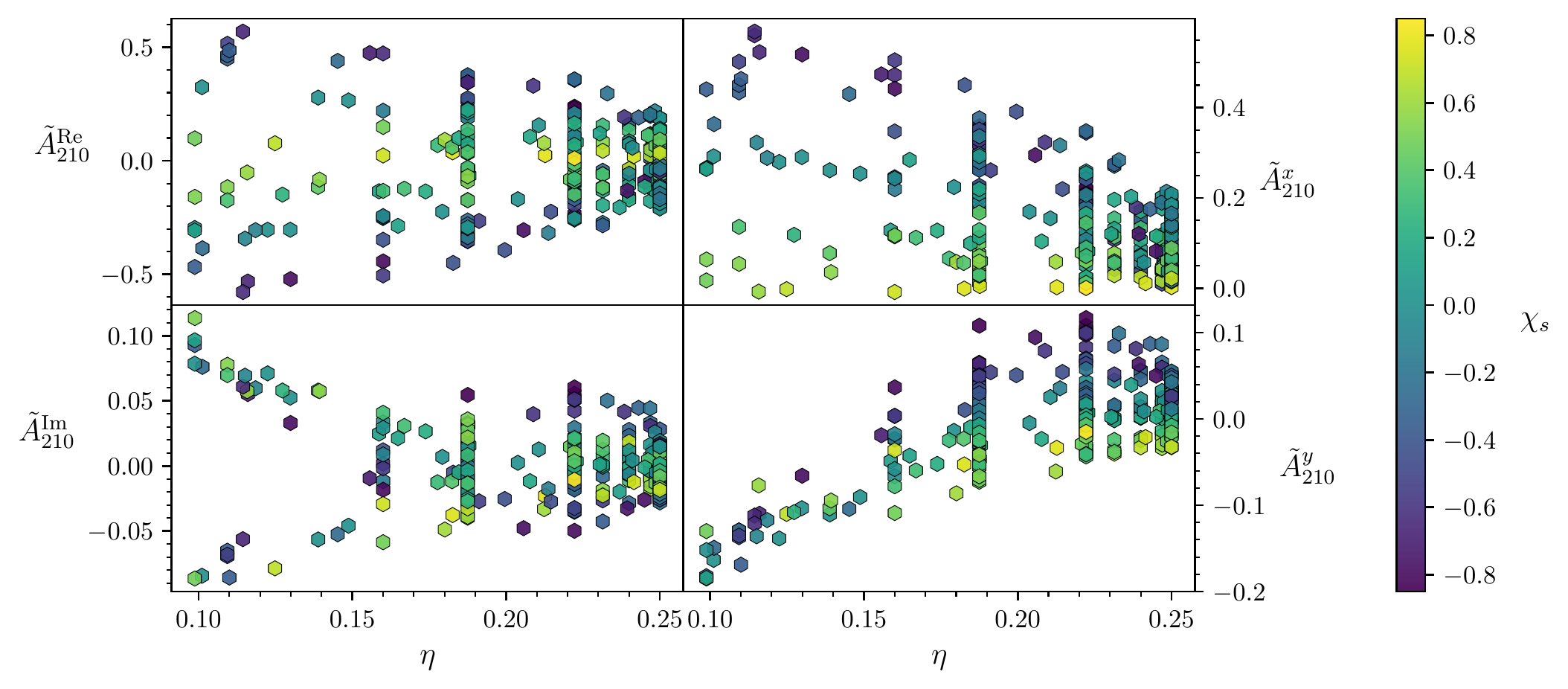}
    \caption{Comparison between the real/imaginary parts of the amplitudes $\tilde A_{210}^{\rm Re}+i\tilde A_{210}^{\rm Im}=A_{210}^{\rm R}e^{i\tilde\phi_{210}}$ (left) and the effective Cartesian components $\tilde A_{210}^{x}+i\tilde A_{210}^{y}=A_{210}^{\rm R}e^{i\beta\tilde\phi_{210}}$ (right). In all subplots, we only plot the data points marked as \textit{inliers}, as defined in Sec.~\ref{sec:training:strategy}.}
    \label{fig:res:5}
\end{figure*}

Figure \ref{fig:res:5} illustrates the advantage of using the Cartesian components over the real and imaginary parts $\tilde A_{lmn}^{\rm Re}+i\tilde A_{lmn}^{\rm Im}=A_{lmn}^{\rm R}e^{i\tilde\phi_{210}}$, when dealing with odd-$m$ modes. For the particular choice $(l,m,n)=(2,1,0)$ we see that, due to the periodicity ${\rm Per}(\tilde\phi)=\pi$, the real and imaginary components form ambiguous patterns with respect to the input parameters --- e.g., for the same values of $\eta$ and $\chi_s$, both positive and negative values of $\tilde A_{210}^{\rm Re,Im}$ exist. On the other hand, the use of Cartesian components $\tilde A_{210}^{x,y}$ ensures that we fit for well behaved quantities.
\subsection{Training strategy}
\label{sec:training:strategy}
In summary, we train GPR fits that map the independent variables $X=\{\delta,\chi_s,\chi_a\}$ to the dependent variables $Y=(\tilde A_{lmn}^x,\tilde A_{lmn}^y)$. For $(l,m,n)\neq(2,2,0)$ the output dimension is 2 (i.e., we do \textit{not} fit $\tilde A_{lmn}^x$ and $\tilde A_{lmn}^y$ separately), while for $(l,m,n)=(2,2,0)$ we just fit $A_{220}$. We use the \texttt{GaussianProcessRegressor} routine implemented within the \texttt{scikit-learn} library  \cite{scikit-learn}. We adopt a radial basis function kernel multiplied by a constant kernel in addition to a white-noise kernel to capture the scattering due to noisy simulations. Prior to fitting, we preprocess the data such that the input and the output are standardized to have zero mean and unit variance. As in Ref.~\cite{Varma:2018aht}, we first subtract a linear fit to the data and then train the GPR on the residuals. 

Since simulations differ in numerical resolution and in the accuracy of the waveform extrapolation, we adopt some mitigation strategies to avoid biasing the regression by NR data with lower quality:
\begin{itemize}
    \item We exclude simulations with $\mathcal{M}_{lm}>0.005$. Then, we penalize the remaining simulations by the inverse squares of their mismatches, i.e.~we perform weighted fits with weights $w_i=1/\mathcal{M}_{i,lm}^2$. We also add terms $\alpha_i=\mathcal{M}_{i,lm}^2$ to the diagonal terms of the kernel matrix as a proxy for additional variances due to the inaccuracy of NR data. Since we fit for variables $A^{\rm R}_{lmn}$ and $\tilde\phi_{lmn}$ that are defined relative to the $(2,2,0)$ modes, uncertainties in the NR recovery of the latter propagate to all other modes. To model for such a propagation, we replace the factors $\mathcal{M}_{li,m}$ with the geometric means $\tilde{\mathcal{M}}_{i,lm}=\sqrt{\mathcal{M}_{i,lm}^2+\mathcal{M}_{i,22}^2}$.
    \item We also implement an outlier detection scheme to further exclude from the final training set the data points that deviate too much from the corresponding GPR predictions: inspired by Ref.~\cite{Forteza:2022tgq}, we first train the GPR on the whole dataset; then we evaluate the residuals $r^2_i$ as defined by Eq.~\eqref{eq:score:abs}; finally, we exclude as outliers all data points with residuals 3 times larger than the median residual.
\end{itemize}
These mitigation procedures result in a restricted set of \textit{inliers} that are eventually used for training. Note that they imply that different modes are trained on a different number of simulations.

We assess the goodness of the GPR training via a tenfold cross validation. Specifically, we split the whole set into 10 subsets and repeat the training ten times: each time, we exclude a different subset from the training set and promote it to a validation set onto which evaluate scoring metrics. The distribution of the scores across the 10 repetitions informs us on the average goodness of the training. After cross validation, the final model is trained on the whole set.
\section{Results}
\label{sec:results}

\subsection{Fit performance}
\label{sec:fit:performance}
In Table~\ref{tab:res:scores} we list the results of our tenfold cross validation for all targeted modes. We report the root mean squared (RMS) score ${\rm RMS}=\sqrt{\sum_{i=1}^N r_i^2/N}$, where the individual residuals $r_i^2$ are given by 
\begin{equation}
    \label{eq:score:abs}
    r^2_i = \frac{\left|\tilde A_{i,lmn}^{x,{\rm GPR}}-\tilde A_{i,lmn}^{x,{\rm fit}}\right|^2
    +\left|\tilde A_{i,lmn}^{y,{\rm GPR}}-\tilde A_{i,lmn}^{y,{\rm fit}}\right|^2}{2}\,.
\end{equation}

We also provide $\text{RMS}_{\sigma}=\sqrt{\sum_{i=1}^N r_{i,\sigma}^2/N}$ score, where $r_{i,\sigma}^2$ are the residuals weighted by the GPR standard deviations given by
\begin{equation}
    \label{eq:score:rel}
    r^2_{i,\sigma} = \frac{\left|\tilde A_{i,lmn}^{x,{\rm GPR}}-\tilde A_{i,lmn}^{x,{\rm fit}}\right|^2}{2\left(\sigma^{x,{\rm GPR}}_{i,lmn}\right)^2}
    +\frac{\left|\tilde A_{i,lmn}^{y,{\rm GPR}}-\tilde A_{i,lmn}^{y,{\rm fit}}\right|^2}{2\left(\sigma^{y,{\rm GPR}}_{i,lmn}\right)^2}\,.
\end{equation}

For each of these two scores, we report the mean and the standard deviation over the 10 test sets used during the cross validation.

The RMS scores estimate the average distance $\Delta_{lmn}$ between the mean GPR predictions for $\tilde{A}^x_{lmn}$ and $\tilde{A}^y_{lmn}$, and the corresponding NR data. They can be converted into estimates for the biases in predicting $A_{lmn}^{\rm R}$ and $\tilde\phi_{lmn}$ as
\begin{equation}
    \label{eq:err:AR}
    \Delta_{A_{lmn}^{\rm R}}\sim\Delta_{lmn}
\end{equation}
and
\begin{equation}
    \label{eq:err:phi}
    \Delta_{\tilde\phi_{lmn}}\sim\frac{\Delta_{lmn}}{A_{lmn}^{\rm R}}\text{ rad}\,.
\end{equation}
Considering the typical values of $A_{lmn}^{\rm R}$ and the values reported in Table~\ref{tab:res:scores}, the average systematic errors on $A_{lmn}^{\rm R}$ are of $\mathcal{O}(10^{-3})$ while the errors on $\tilde\phi_{lmn}$ are of $\mathcal{O}(10^{-2})$ radians. On the other hand, the statistical uncertainties from ringdown observations with current GW detectors are of $\mathcal{O}(10^{-1})$ for $A_{lmn}^{\rm R}$ and $\mathcal{O}(1)$ radians for $\tilde\phi_{mn}$~\cite{Forteza:2022tgq}. Therefore, our GPR predictions can be reliably used for ringdown parameter estimation with current GW detectors.

The $\text{RMS}_\sigma$ scores estimate the average distance between the NR data and the mean GPR predictions, in units of GPR standard deviation. The NR data are about 1 GPR standard deviation away from the mean GPR prediction for all modes, with the largest value corresponding to $(4,3,0)$. In fact, the $(4,3)$ harmonics is subtler to model because of its smaller amplitude and Ref.~\cite{Cheung:2023vki} does not even include it among the modes that are stably recovered. The ratios between the RMS scores and the $\text{RMS}_\sigma$ scores provide an estimate of the average GPR standard deviations over $\tilde{A}^x_{lmn}$ and $\tilde{A}^y_{lmn}$. These are of $\mathcal{O}(10^{-2})$ or less, which implies the intrinsic uncertainties of the GPR predictions are subdominant with respect to the statistical uncertainties of current ringdown observations.
\begin{table}[t!]
    \centering
    \begin{tabular}{c|c|c}
        Mode & RMS score & $\text{RMS}_\sigma$ score\\
        \hline \hline
        \multicolumn{3}{c}{Fundamental tones}\\
        \hline
        $(2,2,0)$ & $(4.9\pm1.3)\times10^{-4}$ & $(7.4\pm2.1)\times10^{-1}$\\
        $(2,1,0)$ & $(5.8\pm0.9)\times10^{-3}$ & $(9.7\pm1.6)\times10^{-1}$\\
        $(3,3,0)$ & $(3.8\pm1.2)\times10^{-3}$ & $(6.8\pm2.9)\times10^{-1}$\\
        $(3,2,0)$ & $(1.60\pm0.67)\times10^{-3}$ & $1.24\pm0.54$\\
        $(4,3,0)$ & $(1.36\pm0.57)\times10^{-3}$ & $1.52\pm0.87$\\
        $(4,4,0)$ & $(1.92\pm0.25)\times10^{-3}$ & $(9.6\pm1.3)\times10^{-1}$\\
        $(5,5,0)$ & $(2.00\pm0.56)\times10^{-3}$ & $1.04\pm0.35$\\
        \hline
        \multicolumn{3}{c}{Overtones}\\
        \hline
        $(2,2,1)$ & $(3.00\pm1.00)\times10^{-3}$ & $(5.13\pm1.41)\times10^{-1}$\\
        $(2,1,1)$ & $(4.6\pm1.0)\times10^{-3}$ & $(9.0\pm2.0)\times10^{-1}$\\
        $(3,3,1)$ & $(6.5\pm3.0)\times10^{-3}$ & $(7.4\pm3.1)\times10^{-1}$\\
        $(3,2,1)$ & $(1.76\pm0.55)\times10^{-3}$ & 
        $(9.3\pm2.9)\times10^{-1}$\\
        $(4,3,1)$ & $(1.22\pm0.33)\times10^{-3}$ & $1.15\pm0.29$\\
        $(4,4,1)$ & $(3.23\pm0.46)\times10^{-3}$ & $(7.6\pm1.1)\times10^{-1}$\\
        $(5,5,1)$ & $(2.45\pm0.26)\times10^{-3}$ & $(8.4\pm1.0)\times10^{-1}$\\
        \hline
        \multicolumn{3}{c}{Quadratic modes}\\
        \hline
        $(2,2,0)\times(2,2,0)$ & $(4.15\pm0.80)\times10^{-3}$ & $(7.6\pm1.5)\times10^{-1}$\\
        $(2,2,0)\times(3,3,0)$ & $(4.21\pm0.78)\times10^{-3}$ & $1.07\pm0.19$
    \end{tabular}
    \caption{RMS scores and ${\rm RMS}_\sigma$ scores, as per Eqs.~\eqref{eq:score:abs} and \eqref{eq:score:rel} respectively, for all the ringdown modes modeled by our GPR. For each score, we report the mean and standard deviation over the 10 tests sets from the tenfold cross validation procedure.
    }
    \label{tab:res:scores}
\end{table}

\begin{figure*}[t]
    \centering
    \includegraphics[width=0.9\textwidth]{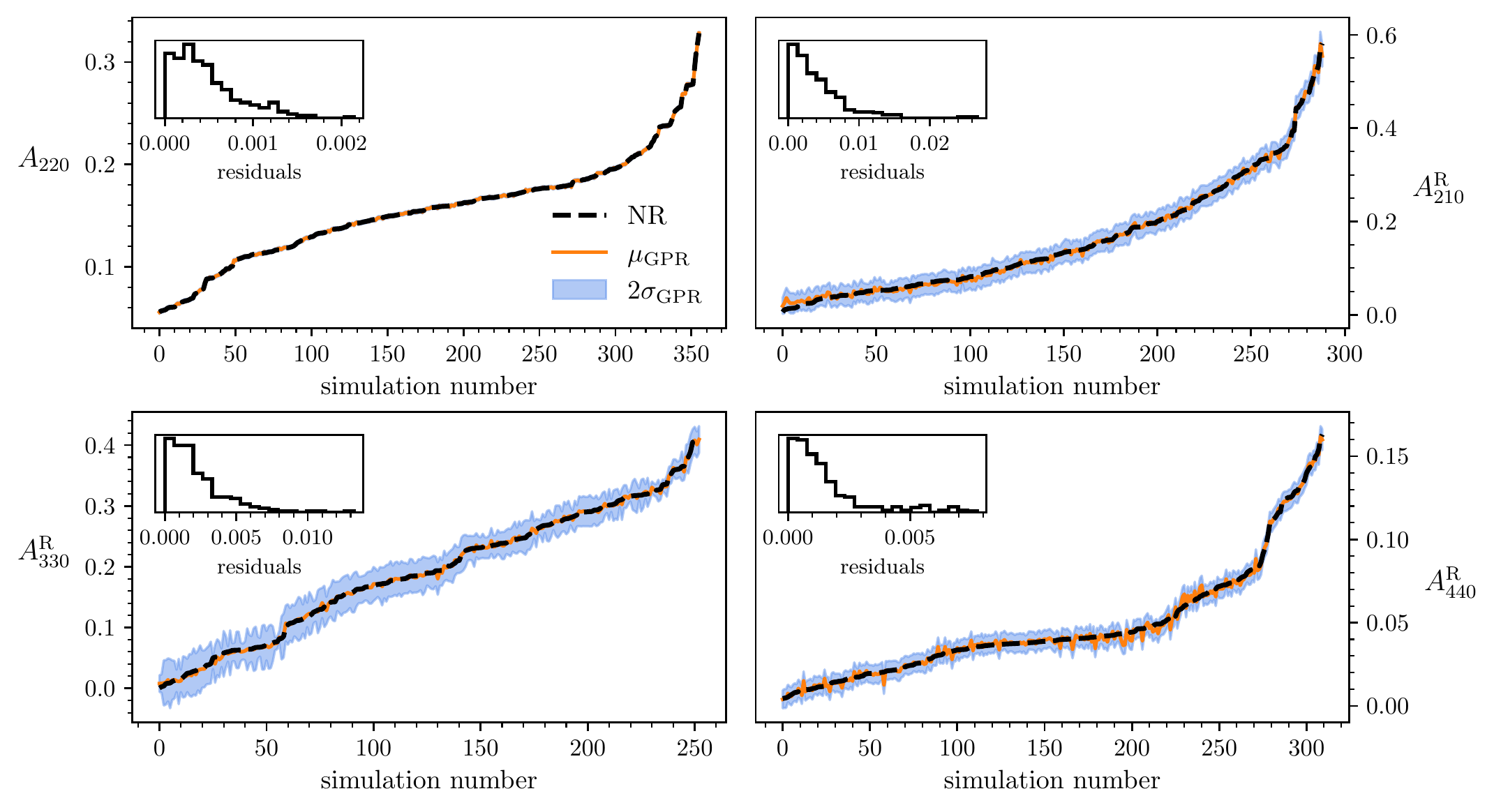}
    \caption{Absolute amplitude $A_{220}$ and relative amplitudes $A_{lmn}^{\rm R}$ for the $(2,1,0)$, $(3,3,0)$ and $(4,4,0)$ modes for all NR simulations included in our training sets. The horizontal axis labels the NR simulations in ascending order of the NR fitted variables $A_{220}^{\rm fit}$ or $A_{lmn}^{\rm R, fit}$. For each of the quantities under consideration, we only include the \textit{inliers} as specified in Sec.~\ref{sec:training:strategy}, which explains the varying length of the horizontal axes. Note that the GPR error band in the upper left panel is too small to be visible. The insets display the distributions of the absolute residuals between the GPR mean predictions and the NR fits.}
    \label{fig:res:1}
\end{figure*}

Figure \ref{fig:res:1} compares NR data against our GPR surrogates, for $A_{220}$ and the amplitude ratios $A_{lmn}^{\rm R}$ of the higher harmonics $(l,m,n)=(2,1,0)$, $(3,3,0)$, and $(4,4,0)$. We show the mean predictions $\mu_{\rm GPR}$ from the GPR together with their $2\sigma_{\rm GPR}$ intervals.\footnote{
We propagate GPR errors over $A^x$ and $A^y$ to $A^{\rm R}$ and $\tilde\phi$ via the chain rules
\[
\sigma_{A^{\rm R}}=\frac{\sqrt{\left(A^x\sigma^x\right)^2+\left(A^y\sigma^y\right)^2}}{A^{\rm R}}
\]
and
\[
\sigma_{\tilde\phi}=\frac{\sqrt{\left(A^y\sigma^x\right)^2+\left(A^x\sigma^y\right)^2}}{\beta\left(A^{\rm R}\right)^2}\text{ rad}\,.
\]
}
NR data follow closely the mean GPR predictions and they are comprised within the $2\sigma_{\rm GPR}$ regions.

\subsection{Comparison with previous studies}
\label{sec:comparison}
\begin{figure*}[t]
    \centering
    \includegraphics[width=0.9\textwidth]{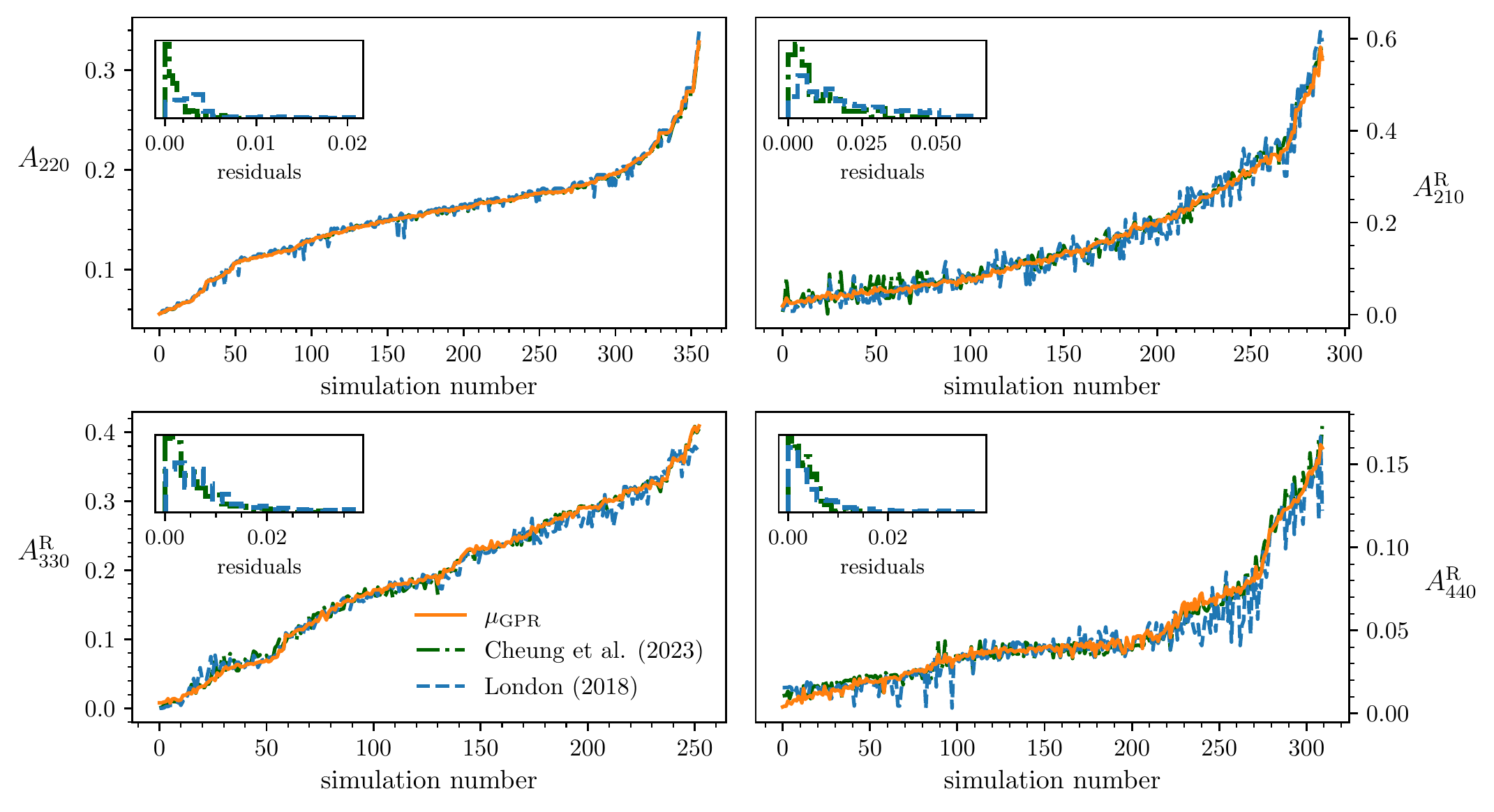}
    \caption{Same as Fig.~\ref{fig:res:1} but showing the comparison between our GPR fits and the closed-form fits from London (2018) \cite{London:2018gaq} and from Cheung et al.~(2023) \cite{Cheung:2023vki}. All fits are evaluated on the NR simulation parameters corresponding to the \textit{inliers} of the respective modes and are plotted in ascending order of NR fitted variables $A_{220}^{\rm fit}$ and $A_{lmn}^{\rm R, fit}$. The insets display the distributions of the absolute residuals between the GPR mean predictions and the corresponding predictions from Refs.~\cite{London:2018gaq,Cheung:2023vki}.}
    \label{fig:res:2}
\end{figure*}

Figure~\ref{fig:res:2} compares our mean GPR predictions agasint existing closed-form fits from the literature, specifically those by \citeauthor{London:2018gaq}~\cite{London:2018gaq} and \citeauthor{Cheung:2023vki}~\cite{Cheung:2023vki}. The former are extracted at $t_{\rm start}=t_{\rm peak}+20M_{\rm i}$ similarly to ours. They are in substantial agreement with our predictions, with differences that can be justified by the use of a different NR catalog, namely the Georgia Tech catalog \cite{Jani:2016wkt} with data up to 2018.

The fits of Ref.~\cite{Cheung:2023vki} are obtained from a subset of 188 nonprecessing simulations from SXS; we thus expect to find much closer numerical agreement. Their results are extracted at varying starting times depending on the simulation and on the stability of the mode under study, which implies there might be discrepancies for those modes that have not yet reached stability at $20M_{\rm i}$ after the peak. However, the fundamental ($n=0$) tones are sufficiently stable after $20M_{\rm i}$ \cite{Cheung:2022rbm}, which can be seen by the fact that the mean GPR predictions and the predictions from Ref.~\cite{Cheung:2023vki} almost overlap throughout the parameter space.
Besides closed-form polynomial fits, Ref.~\cite{Cheung:2023vki} provides alternative interpolation functions constructed via a barycentric interpolator. We obtain similar results when comparing against fits and interpolators: depending on the mode under consideration, we find that fits behave better than interpolators or vice versa. Note, however, that interpolators are not defined outside the convex hull of the calibration region. Here, for concreteness, we report comparisons against the polynomial fits, which is the approach most emphasized in Ref.~\cite{Cheung:2023vki}.
%

\begin{figure}[t]
    \centering
    \includegraphics[width=0.45\textwidth]{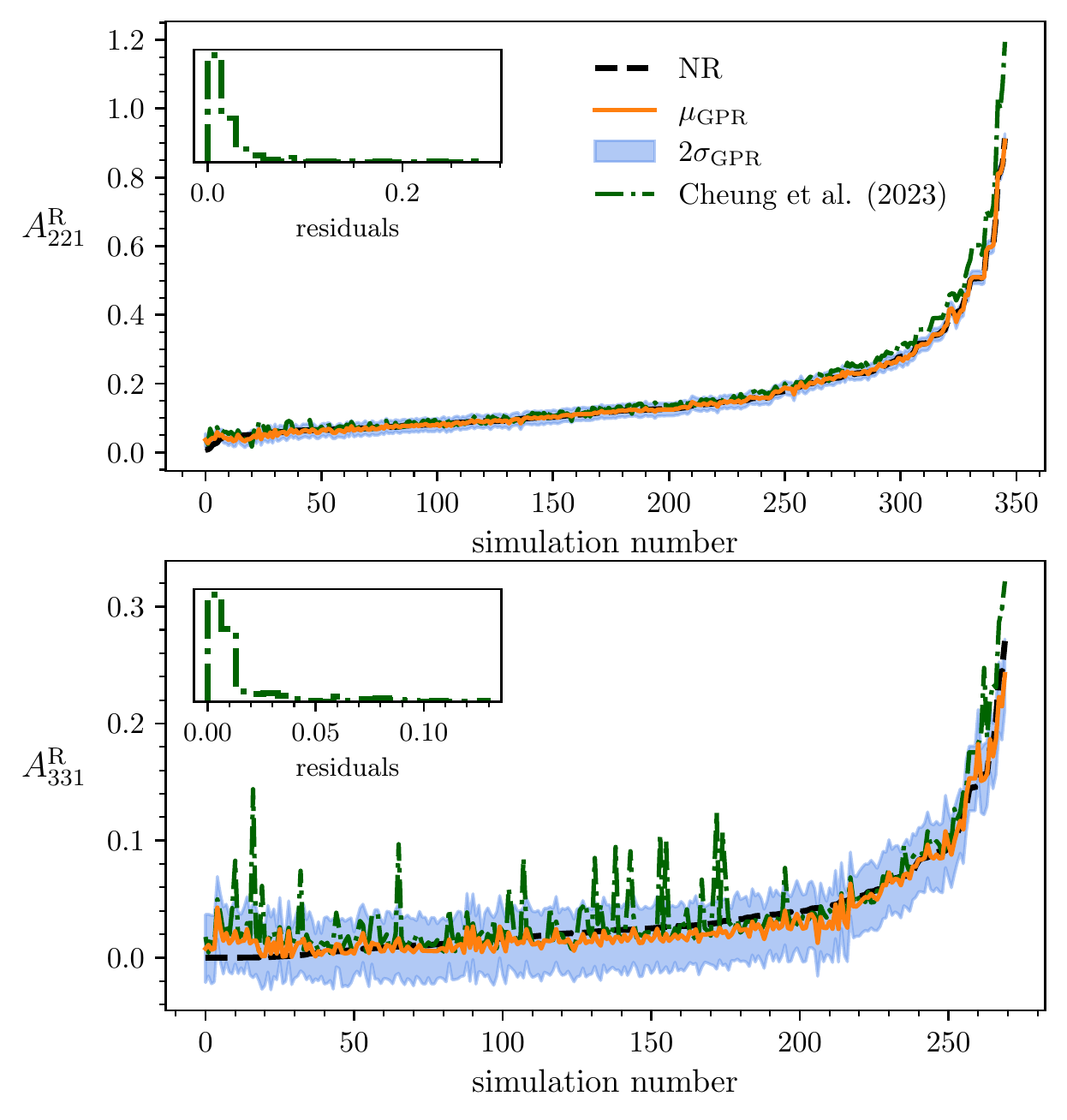}
    \caption{Recovery of the relative amplitudes for the overtones $(2,2,1)$ and $(3,3,3)$. All fits are evaluated on the NR simulation parameters and only including \textit{inliers}. Data are plotted in ascending orders of the NR fitted variables $A_{lmn}^{\rm R, fit}$. The insets display the distributions of the absolute residuals between the GPR mean predictions and the corresponding predictions from Ref.~\cite{Cheung:2023vki}.}
    \label{fig:res:3}
\end{figure}

Figure \ref{fig:res:3} exemplifies our recovery of ringdown overtones. In particular, we show NR data and GPR predictions for $A_{221}^{\rm R}$ and $A_{331}^{\rm R}$ together with predictions from Ref.~\cite{Cheung:2023vki} (Ref.~\cite{London:2018gaq} does not  fit the overtones). For the $(3,3,1)$ mode, the closed-form polynomial fit of Ref.~\cite{Cheung:2023vki} can predict spikes, which are not physical --- enforcing a functional form \textit{a priori} can lead to overfitting outside the calibration region. On the other hand, our nonparametric fit is robust and does not show any such pathology.

\begin{figure}[t]
    \centering
    \includegraphics[width=0.45\textwidth]{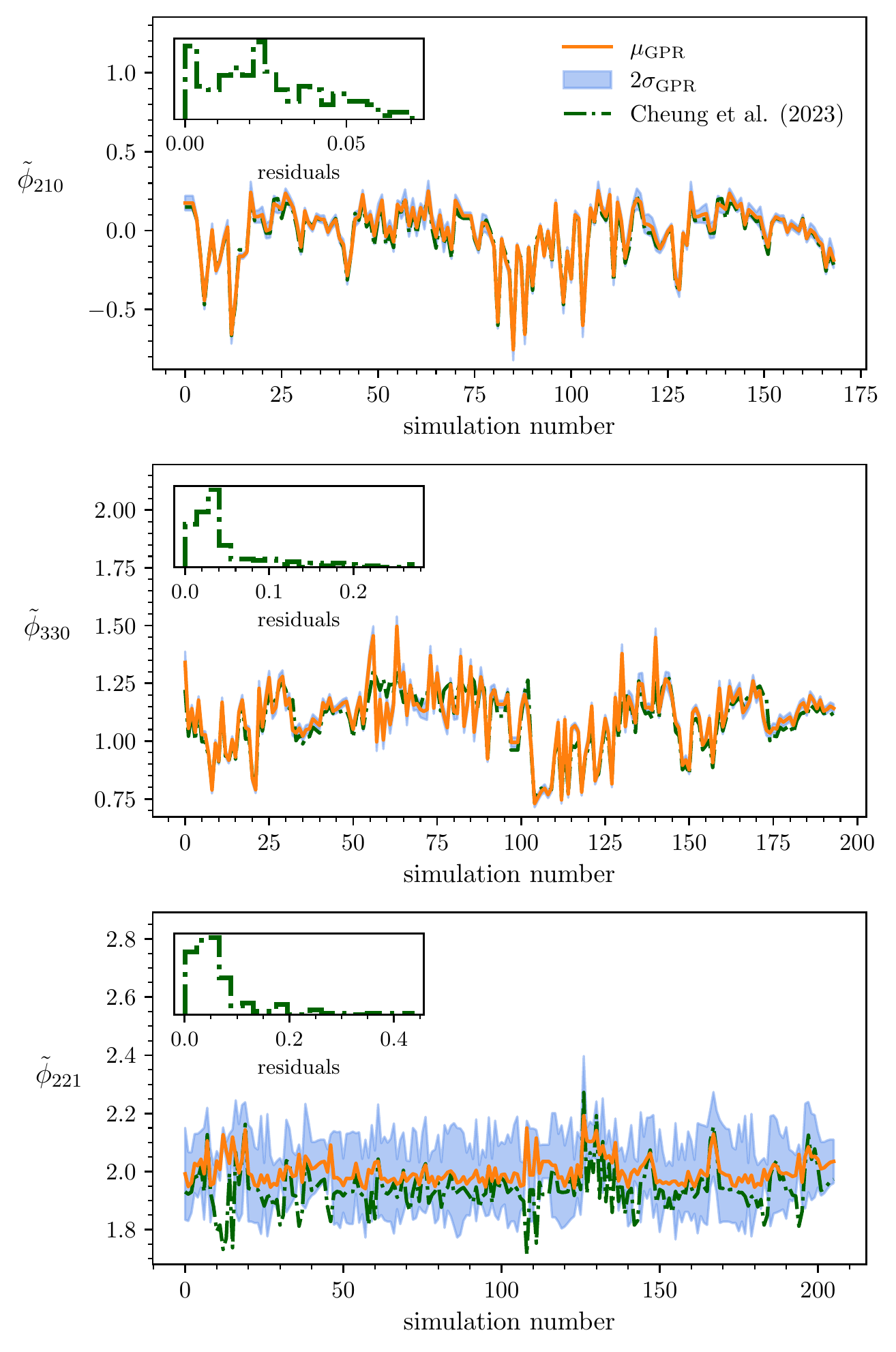}
    \caption{Recovery of the relative phases $\tilde\phi_{lmn}$ for the $(2,1,0)$, $(3,3,0)$ and $(2,2,1)$ modes. All fits are evaluated on NR simulation parameters corresponding to \textit{inliers} and further excluding points with $A_{lmn}^{\rm R, fit}<0.1$. We do not compare our fits against those in \cite{London:2018gaq}, because the latter use a different convention for the definition of the relative phases. The insets display the distributions of the absolute residuals between the GPR mean predictions and the corresponding predictions from Ref.~\cite{Cheung:2023vki}.}
    \label{fig:res:7}
\end{figure}

Figure \ref{fig:res:7} shows GPR predictions for the relative phases of the most excited modes besides $(2,2,0)$, namely $(2,1,0)$, $(3,3,0)$ and $(2,2,1)$. The horizontal axis indicates the NR simulation number; we only consider \textit{inlier} data and we further exclude data with amplitude ratio $A_{lmn}^{\rm R, fit}<0.1$ because phase extractions are less reliable at small amplitudes. We also compare the GPR results with the corresponding predictions from Ref.~\cite{Cheung:2023vki} and show our results are consistent with theirs, at least within the $2\sigma_{\rm GPR}$ error bands.

\subsection{Recovery of the quadratic modes}
\label{sec:recovery}
We comment on the recovery of quadratic $(2,2,0)\times(2,2,0)$ modes within the $(4,4)$ harmonic. This mode has recently drawn attention both from a theoretical standpoint \cite{Mitman:2022qdl,Cheung:2022rbm,Kehagias:2023ctr,Redondo-Yuste:2023seq,Zhu:2024rej,Bucciotti:2024zyp} as well as its detection prospects with next-generation GW detectors \cite{Yi:2024elj}. A relevant parameter here is the ratio $\mathcal{R}_{220\times220}=A_{220\times220}/A_{220}^2$ between the amplitude of the quadratic mode and the amplitudes of the linear modes that contribute to it.

We use our trained GPR to estimate $\mathcal{R}_{220\times220}$. In particular, we generate $10^3$ binary parameters with mass ratios in the range $q\in[1,8]$ and spins in the range $\chi_{z}\in[-1,1]$, and evaluate on them the GPR for the $(2,2,0)\times(2,2,0)$ and for the $(2,2,0)$ modes. After converting the GPR predictions to the absolute amplitudes $A_{220\times220}$ and $A_{220}$, we fit for $\mathcal{R}_{220\times220}$ by a least squares optimization of $|A_{220\times220}-\mathcal{R}_{220\times220}A_{220}^2|^2$. Our best fit result reads
\begin{equation}
    \label{eq:quadratic:ratio:1}
    \mathcal{R}_{220\times220}=0.15\pm0.01
\end{equation}
where the mean and uncertainty are computed over ten repetitions of the estimation procedure. 

We repeat the same exercise for the ratio $\mathcal{R}_{220\times330}$ of the amplitude of the $(2,2,0)\times(3,3,0)$ quadratic modes with respect to $A_{220}A_{330}$ and obtain
\begin{equation}
    \label{eq:quadratic:ratio:2}
    \mathcal{R}_{220\times330}=0.36\pm0.01\,.
\end{equation}

Our results for the amplitude ratios are in good agreement with the analytic perturbation-theory results of Ref.~\cite{Bucciotti:2024zyp}.
\section{Conclusions}
\label{sec:conclusions}
The ringdown phase of GW signals from merging compact binaries contains precious information on the nature of BHs and the underlying theory of gravity. Extracting this ``precision physics'' requires accurate modeling of the QNMs that are excited following BH mergers. In this paper, we have presented a state-of-the-art regression of the QNM amplitudes and phases covering 16 emission modes including overtones and quadratic contributions. Our surrogate is based on GPR, which, at variance with past fits, does not require to assume explicit functional form.

Crucially, our results are restricted to BH binaries with aligned or antialigned spins on quasicircular orbits. In order to reduce systematic biases when analyzing ringdown data, it is essential to extend the existing models to eccentric and/or to precessing binaries. However, extending past studies to these regimes is challenging: it requires to enlarge the number of terms included in parametrized expressions, to propose \textit{ad hoc} phenomenological functional ansatzes and to identify suitable changes of variables in the parameter space that simplify the problem. For example, Ref.~\cite{Carullo:2024smg} recently proposed a model for the ringdown amplitudes of nonspinning eccentric BH binaries with moderate mass ratios, which was made possible by the parametrization introduced in \cite{Carullo:2023kvj}. See also \cite{Finch:2021iip,Zhu:2023fnf} for a detailed study of precessional effects in BH ringdowns. On the other hand, Gaussian Processes are flexible nonparametrized approximants, therefore they are much more suitable for capturing the functional dependence of the QNMs in the higher-dimensional parameter space of eccentric and nonprecessing binaries. Efforts to extend our surrogate to precessing systems are ongoing and will be presented elsewhere.

The typical errors of our reconstructed amplitudes against NR is of $\mathcal{O}(10^{-3})$, which is smaller than the typical measurement errors from current GW interferometers \cite{Capano:2021etf,Forteza:2022tgq}. The time to predict a single (complex) amplitude from our trained GPR is $\mathcal{O}(0.001)$ s. Our fits are thus ready to be ported into parameter estimation study, an endeavor that we plan to tackle in future work. Moreover, they can be used to infer the binary parameters in postprocessing from the agnostic measurement of ringdown amplitudes, similarly to \cite{Capano:2021etf}. To facilitate this, our results are publicly released in a publicly available package called \texttt{postmerger} \cite{pacilio_2024_13220424}.

The accuracy achieved in this work is also sufficient to perform parameter estimations with future detectors, both on the ground \cite{Punturo:2010zz,LIGOScientific:2016wof} and in space \cite{2017arXiv170200786A}, as they are expected to achieve typical errors of $\mathcal{O}(10^{-2})$ on the ringdown amplitudes \cite{Bhagwat:2023jwv,Yi:2024elj}. However, exceptionally loud events could reach measurement errors as low as $\sim 10^{-3}$, for which our surrogate would not be sufficiently accurate. Moreover, the reported errors of our fits are restricted to the parameter space spanned by current NR simulations. More simulations in the training catalog are needed to make our machine-learning fits more precise in regions where current fits are not well calibrated. Crucially, GPR returns estimates of the prediction errors that can be used to inform the placement of new NR simulations in regions where the predictions are less confident (see e.g. Ref.~\cite{Ferguson:2022qkz} for related efforts in this direction). 

A key drawback of GPR is the steep computational complexity of $\mathcal{O}(N^3)$ for training and $\mathcal{O}(N^2)$ for evaluation, where $N$ is the number of data points in the training set. Since GPR scales rather poorly with the number of training data points, this is a drawback that will need to be addressed carefully in future implementations, when more simulations are included in the training catalog.

The following software was used in this work: \texttt{matplotlib} \cite{Hunter:2007}, \texttt{numpy} \cite{harris2020array}, \texttt{scikit-learn} \cite{scikit-learn}, \texttt{scipy}~\cite{2020SciPy-NMeth}, \texttt{sxs} \cite{boyle_2024_12588851}, \texttt{jaxqualin} \cite{Cheung:2023vki}. We used tabulated data for the QNMs of Kerr BHs provided at \cite{Berti_repo,Cardoso_repo}. 

\paragraph*{Note added.} While completing this work, we learned that \citeauthor{MaganaZertuche:2024ajz} conducted a similar study~\cite{MaganaZertuche:2024ajz}.
\section*{Acknowledgments}
\paragraph*{Software.} 

We thank Viola De Renzis, 
Lorena Maga\~na Zertuche, Leo Stein, Mark Ho-Yeuk Cheung, Emanuele Berti, and Gregorio Carullo
for discussions.
C.P. and D.G. are supported by ERC Starting Grant No.~945155--GWmining, 
Cariplo Foundation Grant No.~2021-0555, MUR PRIN Grant No.~2022-Z9X4XS, 
and the ICSC National Research Centre funded by NextGenerationEU. 
S.B. is supported by UKRI Stephen Hawking Fellowship No.~EP/W005727. 
D.G. is supported by MSCA Fellowships No.~101064542--StochRewind and No.~101149270--ProtoBH.
Computational work was performed at CINECA with allocations 
through INFN and Bicocca.%

\appendix
\section{Conventions and parametrizations of quasinormal modes}
\label{sec:qnms}
For each choice of $M$ and $\chi$ and each $(l,m,n)$, Teukolsky's equations \cite{Teukolsky:1972my,Teukolsky:1973ha} admit two distinct solutions $\omega_{lmn}^{[p]}$, labeled by a binary index $p=\{-1,+1\}$ 
namely, the ``prograde'' ($p=+1$) and the ``retrograde''  ($p=-1$) modes. We define prograde-vs-retrograde modes using the same conventions of \cite{Cheung:2023vki}. Under these conventions, solutions with $p=-1$ are also sometimes referred to as the ``mirror'' modes.

NR simulations of quasicircular nonprecessing BH binaries show that mirror modes are subdominant with respect to their prograde counterparts \cite{Dhani:2020nik}. Reference~\cite{Dhani:2021vac} shows that they are most excited for $m=1$ and for high mass ratios, but they do not affect significantly the model systematics. Therefore, in the main text we neglect them and use $\omega_{lmn}$ in place of $\omega_{lmn}^{[+1]}$ (but see Ref.~\cite{Cheung:2023vki} for fitting expressions of the amplitudes of mirror modes as functions of the prograde modes).

The QNMs of Kerr BHs provided in \cite{Berti:2005ys} are tabulated in such a way that tables with $m>0$ correspond to prograde modes, while tables with $m<0$ correspond to retrograde modes. Prograde modes with $m<0$ and retrograde modes with $m>0$ can be obtained via the symmetry
\begin{equation}
    \label{eq:qnms:1}
    f_{l-mn}^{[p]} = -f^{[p]}_{lmn}
    \,,\qquad\tau_{l-mn}^{[p]}=\tau_{lmn}^{[p]}\,.
\end{equation}
Equation \eqref{eq:qnms:1} is such that QNM frequencies have sign ${\rm sgn}(f_{lmn}^{[p]})=p~{\rm sgn}(m)$ while damping times are always positive.

\begin{figure*}
    \centering
    \includegraphics[width=0.8\textwidth]{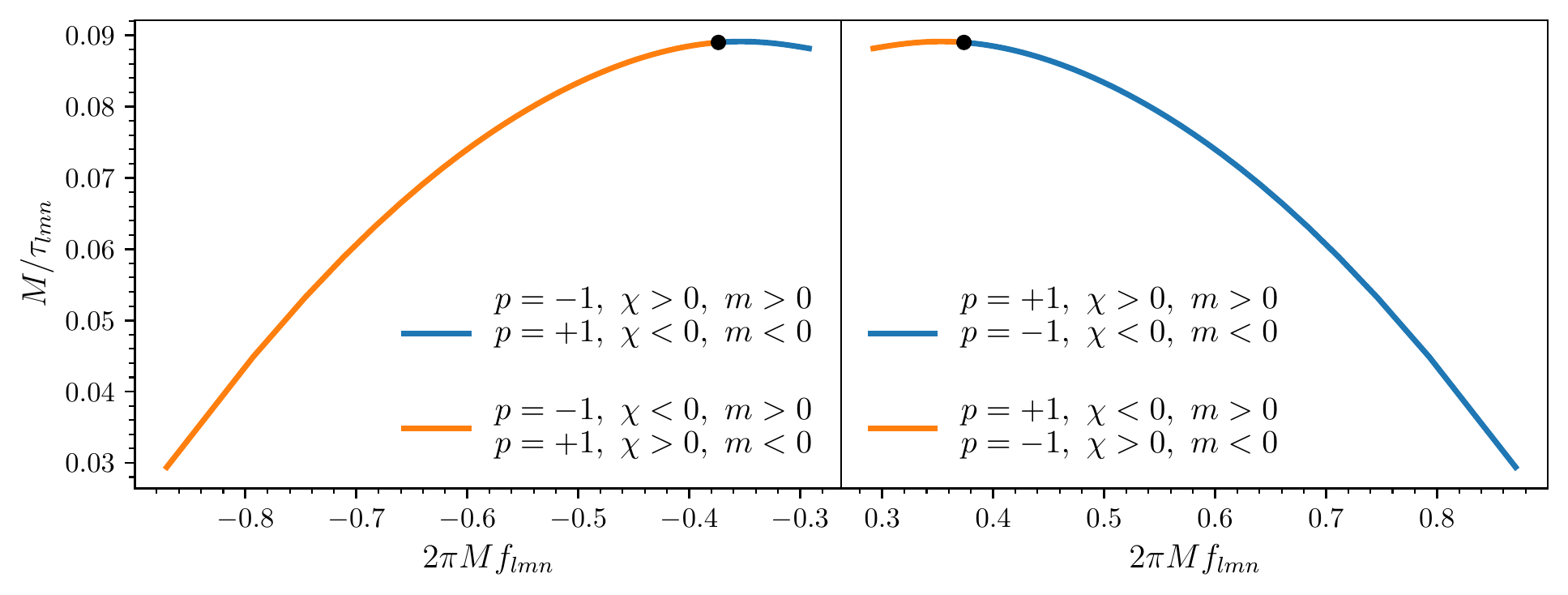}
    \caption{Quasinormal modes of a Kerr black hole with $(l,m,n)=(2,|2|,0)$. \textit{Left:} Retrograde modes with $m>0$ or equivalently prograde modes with $m<0$. \textit{Right:} Prograde modes with $m>0$ or equivalently retrograde modes with $m<0$. Our conventions allow for a continuous treatment of positive and negative spins. The black dots indicate the points where $\chi=0$. Modes on the left panel are the ``mirror'' ones of those on the right panel.}
    \label{fig:qnms}
\end{figure*}

Moreover, data from Ref.~\cite{Berti:2005ys} are tabulated only for positive spins. QNMs for negative spins can be obtained by the symmetry
\begin{equation}
    \label{eq:qnms:2}
    f_{lmn}^{[p]}(-\chi) = f^{[-p]}_{l-mn}(\chi)\,,\qquad\tau_{lmn}^{[p]}(-\chi)=\tau_{l-mn}^{[-p]}(\chi)\,,
\end{equation}
or equivalently
\begin{equation}
    \label{eq:qnms:3}
    f_{lmn}^{[p]}(-\chi) = -f^{[-p]}_{lmn}(\chi)\,,\qquad\tau_{lmn}^{[p]}(-\chi)=\tau_{lmn}^{[-p]}(\chi)\,.
\end{equation}
In the main text, we use Eq.~\eqref{eq:qnms:3} to define the prograde QNMs for configurations with the final spin pointing downward. Figure \ref{fig:qnms} summarizes our conventions for prograde-vs-retrograde modes and for mirror modes.  

Note that alternative parametrizations exist, such as the fits in Ref.~\cite{London:2018nxs}, which give only prograde QNMs for $m>0$ but are otherwise valid for all (positive and negative) spins; all other cases ($p=-1$ or $m<0$) can be reconstructed from Eqs.~ \eqref{eq:qnms:1} and \eqref{eq:qnms:3}. Finally, one can resort to high precision QNM solvers such as that of Ref.~\cite{Stein:2019mop} which also gives mixing coefficients compatible with the conventions used by SXS.
\bibliography{gprringdownaligned}
\end{document}